\documentclass[pre,twocolumn,aps,floats,
floatfix,showpacs,showkeys,amsmath,amssymb,nofootinbib]{revtex4}

\usepackage{dcolumn}
\usepackage{bm}
\usepackage{txfonts}
\usepackage{pxfonts}
\usepackage{graphicx}
\usepackage{psfrag}
\usepackage{times}

\usepackage{epsfig}

\begin{document}

\title{Regularities in stock markets}

\author{Abhin Kakkad}
\email{abhinkakkad@gmail.com}

\author{Harsh Vasoya}
\email{harshvasoya008@gmail.com}

\author{Arnab K. Ray}
\email{arnab_kumar@daiict.ac.in}
\affiliation{Dhirubhai Ambani Institute of Information and 
Communication Technology, Gandhinagar 382007, Gujarat, India}

\begin{abstract}
From the stock markets of 
six countries with high GDP, we study the stock indices,  
{\it S\&P 500 (NYSE, USA)}, 
{\it SSE Composite (SSE, China)}, {\it Nikkei (TSE, Japan)}, 
{\it DAX (FSE, Germany)}, 
{\it FTSE 100 (LSE, Britain)} and {\it NIFTY (NSE, India)}. 
The daily mean growth of the stock values is 
exponential. The daily price fluctuations about the mean 
growth are Gaussian, but with a non-zero asymptotic convergence. 
The growth of the monthly average of stock
values is statistically self-similar to their daily growth.
The monthly fluctuations of the price follow a 
Wiener process, with a decline of the volatility. 
The mean growth of the daily volume of trade is exponential. 
These observations are globally applicable  
and underline regularities across global stock markets. 
\end{abstract}


\maketitle

\section{Introduction}
\label{sec1} 
The main concerns of a stock market analyst are the 
percentage growth rate of 
stock values and the susceptibility of stock prices to 
fluctuations. Speculation in the stock market is encouraged
by the optimism that markets grow over time scales that 
are long enough to average out the impact of adverse fluctuations 
in stock values. Ever since~\citet{bach} proposed his 
theory of speculation, many formal approaches have afforded 
continuous refinement in the analysis of financial 
data (see~\cite{manstan,sccc} and all references therein for a 
historical development of the subject). One object of such researches 
is to discern universal features in stock data~\cite{ps08}, 
whose statistical 
properties do not vary overmuch, irrespective of the market. Financial 
data of world markets do exhibit certain features that appear to 
be broadly common, and as such, markets 
attain some 
predictability. Understanding these familiar patterns 
inspires greater confidence in the speculative activity.  

Our work here has a similar purpose. We study stock indices from the 
stock markets of six countries that are in the top ten of the global ranking 
of GDP. They are USA, China, Japan, Germany, Britain and India. 
Our chosen stock indices are 
{\it S\&P 500 (NYSE, USA)}~\cite{snp}, 
{\it SSE Composite (SSE, China)}~\cite{sse}, 
{\it Nikkei (TSE, Japan)}~\cite{nikkei}, 
{\it DAX (FSE, Germany)}~\cite{dax}, 
{\it FTSE 100 (LSE, Britain)}~\cite{ftse} and 
{\it NIFTY (NSE, India)}~\cite{nifty}.  
We examine stock indices rather than individual stocks, 
because stock indices cover multiple companies across a wide range 
of sectors. Thus, stock indices reflect the overall condition 
of markets more comprehensively than would an individual 
stock~\cite{gpams99,pgams99,ps08}. 
The health of a market can 
be gauged fairly
through the daily price movement of its stock indices, 
the fluctuations they undergo, and the daily volume of stocks traded. 
For these variables, we arrive at well-restricted ranges of numerical 
values
across the markets we have studied. We see that the mean
daily growth
of the value of stock indices is exponential (Sec.~\ref{sec2}). 
The daily fluctuations of a stock index about its exponentially 
growing mean value has a Gaussian distribution, but with 
a non-zero asymptotic convergence for large 
fluctuations (Sec.~\ref{sec3}). 
The growth of the average monthly value of a stock index is 
statistically self-similar to its daily growth (Sec.~\ref{sec4}). 
The monthly variance of the price follows a Wiener
process~\cite{manstan,hull}, with the volatility reducing 
progressively in time (Sec.~\ref{sec4}). The mean growth of the 
number of daily transaction of stocks 
is exponential (Sec.~\ref{sec5}). We present our results graphically 
using the data of {\it NIFTY (NSE, India)}~\cite{nifty}, 
from January, 1997
to April, 2019. 
Similar results from five other stock 
markets~\cite{snp,sse,nikkei,dax,ftse} are 
summarized in Table~\ref{t1}. 
The conclusions derived 
from our analysis of financial data of stock markets of 
six high-GDP countries are globally valid. 

\section{The daily growth of mean stock prices}
\label{sec2} 
\begin{figure}[t]
\begin{center}
\includegraphics[scale=0.65, angle=0]{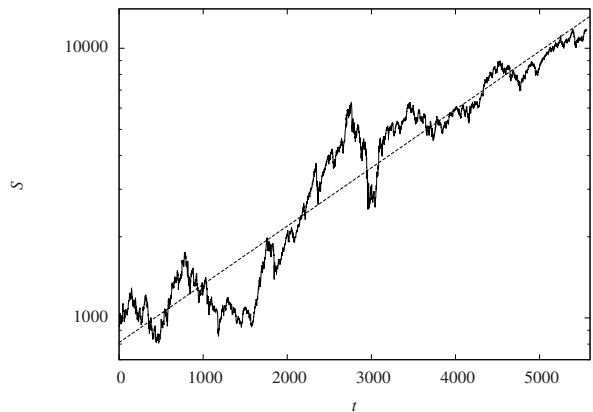}
\caption{\label{f1}\small{The daily mean growth of the 
average price of 
the stock index, {\it NIFTY (NSE, India)}.  
The linear-log plot is modelled by Eq.~(\ref{stock}). Here 
time, $t$, is measured in days, over more than two decades. 
The straight line in this linear-log plot is fitted by the 
least-squares method, and indicates that the mean 
growth of $S$ 
is exponential. In Eq.~(\ref{stock}), with $b=0$, the mean
relative growth rate of stock values is $a$. 
For this plot, $a = 0.05$\% per day. 
Values of $a$ for all the stock indices are provided in 
the second column of Table~\ref{t1}.}}
\end{center}
\end{figure}

The forward relative change of a stock price, $S$, in a finite 
time interval, 
$\Delta t$, is given by~\cite{manstan,hull} 
\begin{equation} 
\label{stock} 
\frac{\Delta S}{S} = a\, \Delta t + b\, \Delta W, 
\end{equation} 
in which $\Delta W$ expresses a Wiener process~\cite{manstan,hull} 
about a background exponential growth of $S$, 
implied by the first term on the right hand side
of Eq.~(\ref{stock}). Under an idealized volatility-free condition 
we set $b=0$ in Eq.~(\ref{stock}), and then integrate it in continuous 
time to get a steady compounded growth of $S$. The integral 
solution of $S$ is exponential in time, $S=S_0 \exp (at)$. 
We, however, express this volatility-free equation slightly 
differently as $\Delta (\ln S)=a\, \Delta t$, and fit it with  
the data of the stock index, {\it NIFTY (NSE, India)}. The fit 
is shown in Fig.\ref{f1}, which is a linear-log plot of the
movement of the daily average  
price of {\it NIFTY (NSE, India)} over more than two decades, 
from January, 1997 to April, 2019. The mean growth of $\ln S$, 
fitted by the least-squares method, is linear in Fig.\ref{f1}, 
demonstrating thereby that the daily mean growth of $S$ occurs
exponentially. 

The daily movement of five other stock indices, 
i.e. {\it S\&P 500 (NYSE, USA)}, {\it SSE Composite (SSE, China)}, 
{\it Nikkei (TSE, Japan)}, {\it DAX (FSE, Germany)} and 
{\it FTSE 100 (LSE, Britain)}, displays the same graphical trend 
as in Fig.\ref{f1}, over time scales of two decades. In all cases, 
the mean relative growth rate of $S$ is $a$, whose values are 
provided in Table~\ref{t1}. Globally, $a$ is confined within  
a range of $0.01$\% to $0.05$\% per day, which seems to be 
surprisingly narrow and precise. 

\begin{table*}
\caption{\label{t1} Summary of the analysis of financial data 
taken from six stock markets, whose names are listed  
in the first column. Regularities across global markets 
are discernible through the numerical values of the  
parameters catalogued below.} 
\begin{ruledtabular}
\begin{tabular}{ccccccc}
Stock Index 
& $a$ (\% per day) & $\mu$ & $\sigma$ & $m$ (per month) & $w$ (per 
month) & $\nu$ (\% per day) \\ \hline 

{\it S\&P 500 (NYSE, USA)} 
& $0.03$  & $0.032$ & $1.203$ & $0.006$ & 
$-2.78 \times 10^{-6}$ & $-$ \\

{\it SSE Composite (SSE, China)} 
& $0.04$ & $0.101$ & $2.805$ & $0.009$ & 
$-3.09 \times 10^{-5}$ & $0.12$ \\

{\it Nikkei (TSE, Japan)} 
& $0.01$ & $0.027$ & $1.494$ & $0.003$ & 
$-1.02 \times 10^{-6}$ & $0.53$ \\

{\it DAX (FSE, Germany)} 
& $0.03$ & $0.025$ & $1.466$ & $0.006$ & 
$-4.42 \times 10^{-6}$ & 
$0.02$ \\

{\it FTSE (LSE, Britain)} 
& $0.01$ & $0.011$ & $1.165$ & $0.002$ & 
$-1.78 \times 10^{-6}$ & $\,\,\,0.003$ \\

{\it NIFTY (NSE, India)}  
& $0.05$ & $0.057$ & $1.495$ & $0.010$ & 
$-3.41 \times 10^{-6}$ & $0.04$ \\
\end{tabular}
\end{ruledtabular}
\end{table*}

\section{Gaussian fluctuations in stock prices}
\label{sec3}
\begin{figure}[t]
\begin{center}
\includegraphics[scale=0.65, angle=0]{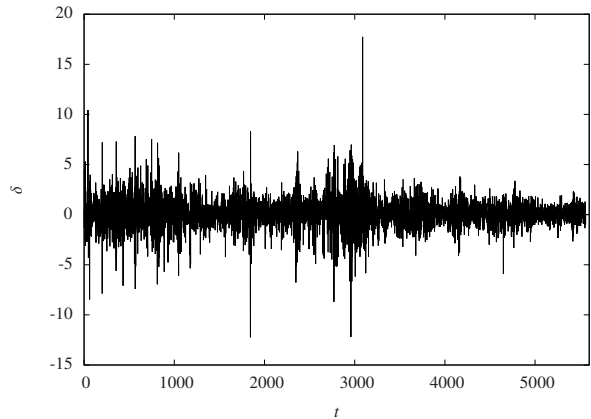}
\caption{\label{f2}\small{The time series of the daily 
percentage fluctuation of prices in the stock index, 
{\it NIFTY (NSE, India)}. 
As in Fig.\ref{f1}, the time, $t$,
is measured in days. The daily percentage 
fluctuation of prices is quantified by $\delta$, which, over 
two decades, has an equal distribution of positive 
and negative values about $\delta =0$. Most of the fluctuations 
are small, but a very few fluctuations have large values in both
extremes. Taken together, all these features are captured by 
Eq.~(\ref{gauss}), which accommodates the possibility of the 
rare but large fluctuations.}}
\end{center}
\end{figure}

\begin{figure}[t]
\begin{center}
\includegraphics[scale=0.65, angle=0]{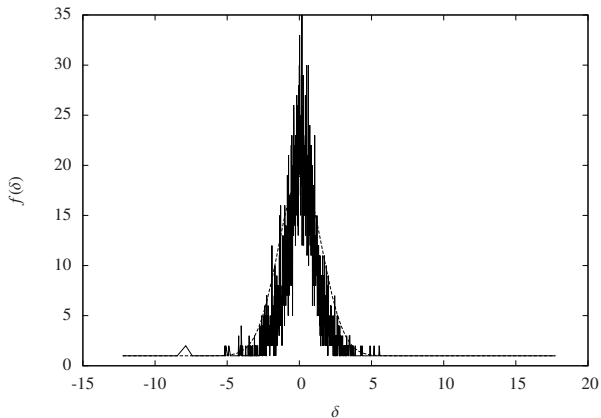}
\caption{\label{f3}\small{The unnormalized frequency distribution 
of the daily percentage fluctuation of prices in the 
stock index, {\it NIFTY (NSE, India)}. 
The distribution appears Gaussian, and is centred 
around a mean value, $\mu = 0.057$, with a standard deviation, 
$\sigma = 1.495$. For large fluctuations, the asymptotic 
convergence is 
$f(\delta) \longrightarrow 1$, as Eq.~(\ref{gauss}) confirms. 
Values of $\mu$ and $\sigma$ for all the stock indices 
are to be found, respectively, in the third and fourth columns 
of Table~\ref{t1}.}}
\end{center}
\end{figure}

By setting $b=0$ in Eq.~(\ref{stock}), we have so far ignored the 
fluctuations about the mean exponential growth of a stock price, $S$. 
In reality, 
however, as we can see very clearly in Fig.\ref{f1}, stock prices 
fluctuate noticeably about the mean exponential growth 
(represented by the straight line in Fig.\ref{f1}). To quantify 
the fluctuations, we 
define a new variable, $\delta$, which is the daily percentage 
change that a stock index undergoes, with respect to the 
value of the previous day. Positive fluctuations are given by 
$\delta >0$, and negative fluctuations by $\delta <0$. 
The time series of such fluctuations, for the stock index, 
{\it NIFTY (NSE, India)}, is plotted in Fig.\ref{f2}, which shows
that positive and negative fluctuations are balanced about 
$\delta =0$. Most of the fluctuations are close to  
$\delta =0$, with a very few cast far away. The 
unnormalized frequency 
distribution of these fluctuations has a Gaussian profile in 
Fig.\ref{f3}, but with an asymptotic convergence
towards unity. To model all these aspects, we
put forward a Gaussian function, with unity added to it, as     
\begin{equation} 
\label{gauss} 
f(\delta)=1+f_0 \exp \left[-\frac{(\delta -\mu)^2}{2\sigma^2}\right],  
\end{equation} 
in which the mean of the frequency distribution, 
$\mu =\langle \delta \rangle$, and the standard deviation, 
$\sigma = \sqrt{\langle \delta^2 \rangle -\mu^2}$. 
The values of both $\mu$ and $\sigma$, calibrated with the 
data of our chosen six stock indices, 
are listed in Table~\ref{t1}. With the frequency distribution 
clustering around $\delta = 0$ in Fig.\ref{f3}, we expect $\mu$ 
to have a small value. Supporting this expectation, 
Table~\ref{t1} shows us that $\mu$ is zero up to the 
first place of decimal for all the stock indices, 
except {\it SSE Composite (SSE, China)}. Likewise, the first 
significant figure of $\sigma$ is unity, except for 
{\it SSE Composite (SSE, China)}.

The theory of unbiased and random fluctuation of stock prices, 
described by a continuous Gaussian function, was 
introduced in financial markets by~\citet{bach}. The randomness
of prices is a response to a ceaseless stream of unpredictable 
and uncorrelated information that enters the market continuously.  
While this process plays out on the microscopic scale 
of individual prices, its macroscopic imprint is manifested 
through a similar randomness in the market indices.  
Fluctuations in financial markets are, therefore, 
a collective macro-outcome of a diversity of microscopic causes 
that are independent of one another. The randomness
agrees theoretically with the Gaussian 
function, but only if the time scale of the fluctuations is  
long enough~\cite{gpams99,pgams99,manstan}, such as 
the scale of a day in our study. On shorter time scales, however, 
market indices exhibit scaling properties~\cite{ms95}, leading 
to an inverse cubic law of stock value 
fluctuations~\cite{gmas98,ps08}. Such fluctuations converge 
more slowly towards an asymptotic limit than the Gaussian function 
does. Slow 
convergence is the hallmark of power laws, but we can
mimic a slow convergence even with a Gaussian function, by 
adding a constant of unity to the right hand side of Eq.~(\ref{gauss}). 
This reconciles a continuous Gaussian function with a discrete
frequency distribution, whose lower limit is practically 
restricted to unity.\footnote{Discrete frequency distributions 
in complex systems are similarly modelled by power laws that 
converge asymptotically to a non-zero limit~\cite{nnr12,nnr14}.} 
Consequently even for large fluctuations, the unnormalized 
frequency distribution will asymptotically converge to unity. 
When $\delta \longrightarrow \infty$, the form of Eq.~(\ref{gauss}),
despite being Gaussian in essence, theoretically 
ensures the convergence, $f(\delta) \longrightarrow 1$.  

Since the Gaussian distribution in Fig.\ref{f3} is based on the daily
data of {\it NIFTY (NSE, India)}, from January, 1997 to April, 2019, 
it is worth noting that nearly two decades ago, 
an exponential distribution of price fluctuations in 
{\it NSE (India)} was reported~\cite{mpss04}. The reason for the 
exponential distribution, which is intermediate between a power-law 
distribution and a Gaussian distribution, was put down to 
conservative tendencies in the evolving Indian financial market, 
and the control of the state over it~\cite{mpss04}. However, in 
the two decades since then, the Indian financial market has matured 
towards greater liberalization. Therefore, we are of the view that 
India, being a country that ranks high in terms of GDP, 
more faithfully reflects at present 
the trends of developed financial markets. Our point of view
is supported by the conclusion of a study that ``far from 
being different, the distribution of price fluctuations in Indian 
markets is quantitatively almost identical to that of 
developed markets."~\cite{sp} 
The Gaussian distribution in Fig.\ref{f3} is not because of
the stage of development of the Indian financial market. 
Rather it is due to the time scale of the price fluctuations in 
our study, for it is recognized well by now that on longer time 
scales, price fluctuations assume the Gaussian 
form~\cite{gpams99,pgams99,manstan,ps07}.

\section{The Wiener process and volatility}
\label{sec4} 
Looking back at Eq.~(\ref{stock}), 
we consider first the basic 
nature of a Wiener process. If a variable, $W$, undergoes a Wiener 
process, then its change, $\Delta W$, in a discrete interval of 
time, $\Delta t$, is  
$\Delta W = \epsilon \sqrt{\Delta t}$~\cite{hull}. 
Here $\epsilon$ has a standardized normal distribution with a
zero mean and a standard deviation of unity~\cite{hull}. 
For different short intervals, $\Delta t$, $\Delta W$ will have 
independent values. Over a time interval, $T$,  
the mean, $\langle \Delta W \rangle =0$, and 
the standard deviation,  
$\sqrt{\langle (\Delta W)^2 \rangle} \sim T^{1/2}$~\cite{hull}. 
\begin{figure}[]
\begin{center}
\includegraphics[scale=0.65, angle=0]{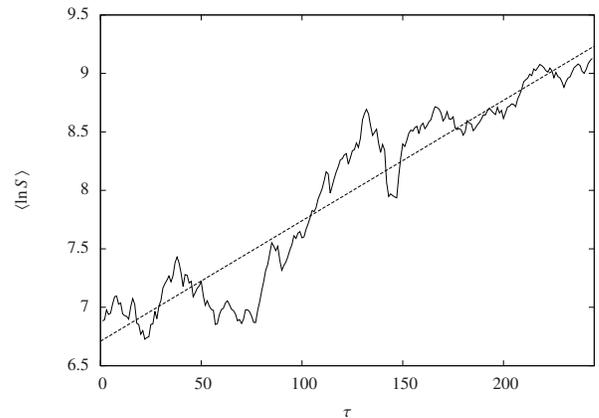}
\caption{\label{f4}\small{The growth of the
monthly average of $\ln S$ for {\it NIFTY (NSE, India)},
as opposed to its daily growth in Fig.\ref{f1}.
The time, $\tau$, is scaled in months, and spans more than two
decades. The straight line, showing the mean growth, is fitted
by the least-squares method, and its slope is 
$m = 0.01$ per month.
Values of $m$ for all the stock indices are in
the fifth column of Table~\ref{t1}. This plot 
looks statistically self-similar to the plot in Fig.\ref{f1}, despite
differing widely in time scales.}}
\end{center}
\end{figure}
\begin{figure}[]
\begin{center}
\includegraphics[scale=0.65, angle=0]{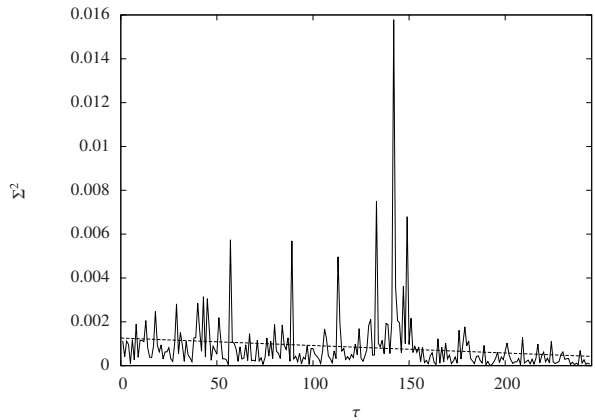}
\caption{\label{f5}\small{The Wiener variance about the
monthly average of $\ln S$ for {\it NIFTY (NSE, India)}
decreases with time, $\tau$ (in months). 
The straight line, fitted by the least-squares method,
traces the mean decline, with a slope of
$w= -3.41 \times 10^{-6}$ per month. With $w <0$, volatility 
also reduces with time. 
Values of $w$ for all the stock indices are in
the sixth column of Table~\ref{t1}. The tallest spike in
this plot coincides with the global economic recession around
the year, 2008, and all stock markets show 
a similar spike about the same year.}}
\end{center}
\end{figure}

Stock price variations follow a generalized Wiener 
process~\cite{hull},  
viewed also as a geometric Brownian 
motion~\cite{manstan,sccc,hull}. The generalized Wiener process 
has a non-zero mean that drifts linearly in time~\cite{hull}, as 
the first term on the right hand side of Eq.~(\ref{stock}) suggests. 
About this linear mean drift, there is a stochastic component, given 
by the second term on the right hand side of Eq.~(\ref{stock}), now
to be read as $\Delta W = \epsilon \sqrt{\Delta t}$. On a daily 
time scale, $t$, the mean growth is depicted by the straight line in 
Fig.\ref{f1}, 
while the stochastic fluctuations (related to the volatility of 
the stock value~\cite{hull}) are seen in the jagged features 
about the straight line. Likewise, all through January, 1997 to 
April, 2019, on a monthly time scale, which 
we write as $\tau$ to distinguish it from $t$, the 
straight line in Fig.\ref{f4} shows the mean growth of the 
monthly average of the natural logarithm of the price of the stock 
index, {\it NIFTY (NSE, India)}. About the mean linear growth of 
$\langle \ln S \rangle$, whose slope is $m$ in Fig.\ref{f4}, the 
jaggedness shows 
the usual stochastic behaviour --- the volatility. 
Notwithstanding the noticeable difference in the time scales,  
i.e. a day for Fig.\ref{f1} and a month for Fig.\ref{f4}, there 
is a statistical self-similarity between the two plots. This is 
a generic characteristic of all the stock indices that we have 
studied. Consistent with this observation, we find from 
Table~\ref{t1} that 
across all stock indices, there is a high correlation coefficient 
of $0.987$ between $a$ and $m$. 

\begin{figure}[t]
\begin{center}
\includegraphics[scale=0.65, angle=0]{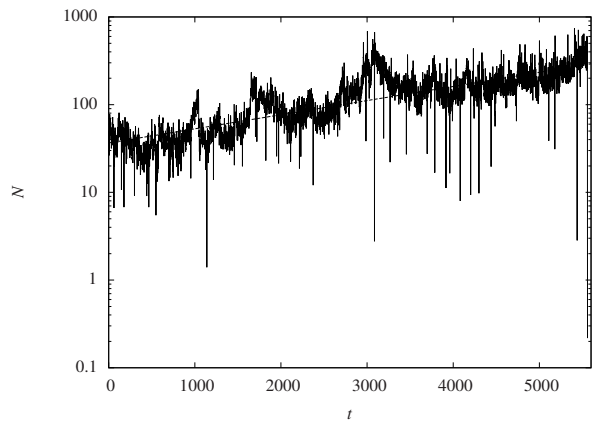}
\caption{\label{f6}\small{The growth of the daily trade volume
of the {\it NIFTY (NSE, India)} index. The number of
daily transactions, $N$, is scaled by $10^6$, and the time,
$t$, is scaled in days.
The straight line in this linear-log plot, fitted by the
least-squares method, implies an exponential mean
growth of $N$. The slope of the straight line,
$\nu = 0.04$\% per day, gives the mean relative growth
rate of the daily volume of trade.
Values of $\nu$ for all the stock indices are in
the last column of Table~\ref{t1}.}}
\end{center}
\end{figure}
From the {\it NIFTY (NSE, India)} index, besides estimating  
$\langle \ln S \rangle$ on a monthly time scale, we also estimate 
the standard deviation of $\ln S$ over the same month. The standard
deviation, $\Sigma$, is related to the volatility of the stock 
price~\cite{hull}. This volatility is clearly revealed by the 
fluctuations 
about the linearly growing mean of $\langle \ln S \rangle$ 
in Fig.\ref{f4}. We also see in Fig.\ref{f4} that the 
fluctuations reduce progressively as both $\tau$ and $S$ 
increase. Hence, the volatility of the stock price subsides 
with the passage of time. We can understand this phenomenon
for a Wiener process 
by recasting Eq.~(\ref{stock}) as 
$(\Delta S/S) = a \Delta t + b \epsilon \sqrt{\Delta t}$, 
on whose right hand side, the first term stands for the mean 
relative growth
of $S$ and the second term stands for the fluctuations about the
mean growth. In an interval of time, $\Delta t$, the mean relative  
growth varies linearly with $\Delta t$, while the fluctuations 
vary as $(\Delta t)^{1/2}$. Thus, on large time scales, the relative
growth of $S$ is dominated by the linear mean growth, against 
which the fluctuations become much weakened. Scaling the time in 
months, this is exactly what we see in Fig.\ref{f4}. Further 
to this point of view, Fig.\ref{f5} demonstrates how the 
variance of the monthly prices, $\Sigma^2$, generally decreases 
with time, $\tau$. A quantitative measure of this decline is found 
in the slope, $w$, of the straight line, $\Sigma^2 = w \tau$, 
fitted with the data in Fig.\ref{f5} by the least-squares method. 
With $w <0$ for the {\it NIFTY (NSE, India)} index, as well as 
for all the other stock indices in Table~\ref{t1},  
stability against volatility is seen 
to be a general tendency of mature stock markets. 

In passing, we note that Fig.\ref{f5} has a prominent spike
in the monthly variance. 
This spike corresponds to the economic recession that disrupted 
global markets around the year, 2008. 
This disruption is universally captured by a very tall spike 
in the monthly 
variance of all the stock indices we have studied.  

\section{The daily volume of trade}
\label{sec5} 
A reliable yardstick of the health of a stock 
market is the number of daily transactions 
in the market~\cite{smdq05,smdq16}. 
The linear-log plot in Fig.\ref{f6} shows 
an exponential mean growth of the trade volume at 
$0.04$\% per day (which is about $15$\% per annum) for 
the {\it NIFTY (NSE, India)} index. 
Volatility about the mean growth is also clearly seen in 
Fig.\ref{f6}.   
Other global markets display the same behaviour, going by the 
relative growth rate of their daily transactions in Table~\ref{t1}. 
These observations qualitatively correspond to established views 
about how trading volume grows with time in financial 
markets~\cite{ps08,smdq05,smdq16}. 

The growth trends in Figs.\ref{f1},~\ref{f4}~\&~\ref{f6} look 
similar, although the first two relate
to price fluctuations (one on a daily time scale and 
the other on a monthly time scale), while the last relates to 
the fluctuations of the daily transactions. This, however, 
is not surprising, because, for better or for worse, 
price fluctuations do affect speculation in 
a financial market, and this in turn has a bearing on its 
volume of trade~\citep{smdq16}. The inter-dependency between 
the two motivates the suggestion that the trading volume can 
be a suitable measure of volatility in financial 
markets (see~\cite{smdq16} and relevant references therein).  

\section{Conclusions}
\label{sec6} 
The indices of the stock markets of six countries with high
GDP show that markets experience long-term growth 
of stock values and the amount of trade. Mature 
markets tend to be stable against volatility. These 
qualitative views
are quantitatively favoured by the numerical values we 
present in Table~\ref{t1}. The closeness of the values of 
each parameter in Table~\ref{t1} satisfies our quest 
for regularities in stock markets. 

\bibliography{kvr062020}
\end{document}